\RequirePackage{rotating}

\documentclass{nws}

\usepackage{epsfig}
\usepackage{times}
\usepackage{algpseudocode}
\usepackage{mathrsfs}
\usepackage{amssymb}
\usepackage{multirow}

\title[Partial Community Merger Algorithm]
{Efficient Detection of Communities with Significant Overlaps in Networks: Partial Community Merger Algorithm}

\author[Elvis H.W. Xu and Pak Ming Hui]
	{Elvis H.W. Xu and Pak Ming Hui\\
	Department of Physics, The Chinese University of Hong Kong,\\
	Shatin, New Territories, Hong Kong, China \\
	\email{\{hwxu,pmhui\}@phy.cuhk.edu.hk}}

\date{\today}

\begin{document}

\maketitle

\begin{abstract}
Detecting communities in large-scale networks is a challenging task when each vertex may belong to multiple communities, as is often the case in social networks. The multiple memberships of vertices and thus the strong overlaps among communities render many detection algorithms invalid. We develop a Partial Community Merger Algorithm (PCMA) for detecting communities with significant overlaps as well as slightly overlapping and disjoint ones. It is a bottom-up approach based on properly reassembling partial information of communities revealed in ego networks of vertices to reconstruct complete communities. Noise control and merger order are the two key issues in implementing this idea. We propose a novel similarity measure between two merged communities that can suppress noise and an efficient algorithm that recursively merges the most similar pair of communities. The validity and accuracy of PCMA is tested against two benchmarks and compared to four existing algorithms. It is the most efficient one with linear complexity and it outperforms the compared algorithms when vertices have multiple memberships. PCMA is applied to two huge online social networks, Friendster and Sina Weibo. Millions of communities are detected and they are of higher qualities than the corresponding metadata groups. We find that the latter should not be regarded as the ground-truth of structural communities. The significant overlapping pattern found in the detected communities confirms the need of new algorithms, such as PCMA, to handle multiple memberships of vertices in social networks.

\medskip
\setlength\parindent{0pt}\textbf{Keywords:} Community Detection, Community Structure, Overlapping Community, Partial Community, Social Networks, Vertex Memberships, Metadata Groups, Linear Complexity
\end{abstract}

\section{Introduction}
\label{sec:introduction}

Community structure is commonly found in networked systems in
nature and society~\cite{Girvan:2002ez}.  While it is almost
common sense to realize the existence of communities, extracting
such mesoscopic structures efficiently and accurately remains a
challenging task and yet it is crucial to the understanding of the
functionality of these systems.  Although the definition of
community remains ambiguous and a commonly accepted definition is
lacking, many detection algorithms have been developed in the past
decade~\cite{Fortunato:2010iw} with most of them motivated by an intuitive notion that there should be more edges within the community than
edges connecting to the outside~\cite{Seidman:1983kf,
Girvan:2002ez, Radicchi:2004br}. This viewpoint on what a
community is about implies that communities are almost disjoint, and it
is behind the design of non-overlapping community detection
algorithms. However, it was soon found by empirical studies that
it is common for communities to overlap, i.e. each vertex may have
multiple memberships~\cite{Palla:2005cj} and thus it may be shared
by communities.  Several approaches have been proposed for
detecting overlapping communities, including clique
percolation~\cite{Palla:2005cj}, link
partitioning~\cite{Evans:2009kg, Evans:2010cf, Ahn:2010dj}, local
expansion and
optimization~\cite{Baumes:2005tf,Lancichinetti:2009dy,Lancichinetti:2011gn},
and label propagation~\cite{Raghavan:2007by, Gregory:2010hm,
Xie:2011hc}.
Some of them inherit the notion of non-overlapping communities and then loose constraint to allow vertices to be shared by communities. For example, algorithms of local expansion and optimization usually use a fitness function that is positively correlated to the ratio between the number of internal and external edges of a community. We remark that the fitness function is fine with disjoint or slightly overlapping communities, but is not applicable to significantly overlapping communities of which most vertices have multiple memberships. For example, if most members of a community have more than one membership, it is very likely that the community has much more edges to the outside than edges within the community, which is often the case in social networks. We think a new concept fundamentally different from non-overlapping communities is needed that imposes no constraints or implications on the fraction of overlapping vertices and the number of communities a vertex may have. Taking the two aspects into consideration, we propose a general concept of overlapping community that each member of a community should connect to a certain fraction of the other members. There are no constraints on whether the community has more internal than external edges or whether a member devotes most of its edges to this community. This intuitive idea is similar to \textit{$k$-core}, but it is fraction-based and we name it \textit{$f$-core}. Unlike $k$-core that cannot overlap, an $f$-core allows its members to belong to arbitrary number of $f$-cores and thus it is suitable to be a definition of overlapping community. Yet, $f$-core only focuses on edge densities and does not specify how members are connected to each other.
To make it a useful concept in the context of
communities, a further constraint that a
community has members who are densely connected to each other and
thus a relatively high value of clustering coefficient will prove
effective.

\begin{figure}[htbp]
   \centering
   \epsfig{file=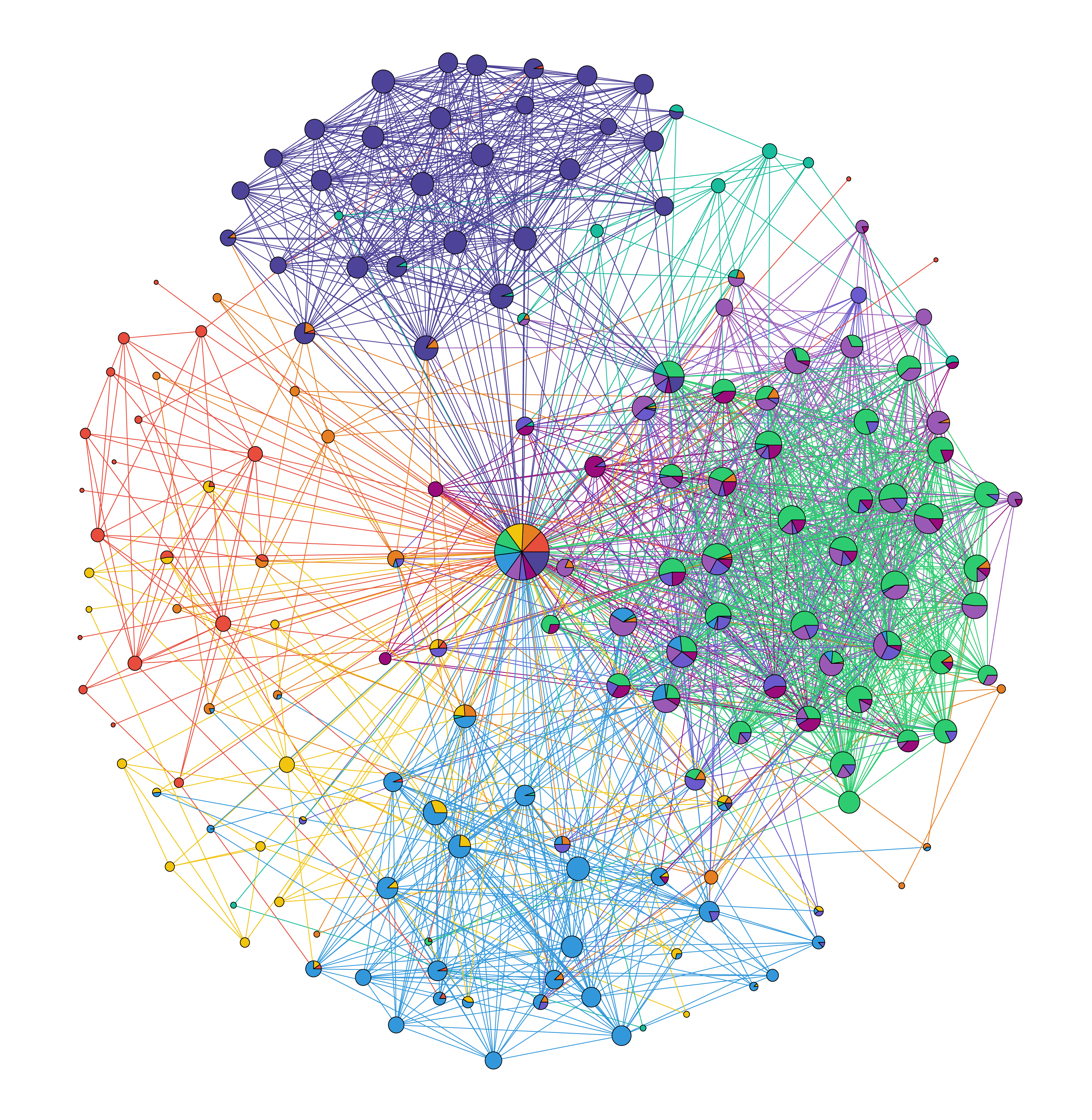, width=0.8\linewidth}
   \caption{(Color online) An ego network of a vertex provides the local information
   and reveals several partial communities.  The network was
   constructed from data collected from Sina Weibo, an online
   social network akin to the hybrid of Facebook and Twitter.  The partial communities
   are found by an existing algorithm as described in Appendix~\ref{sec:appendix_1}. PCMA
   is an efficient and accurate algorithm for detecting
   complete communities in a huge network by
   properly merging partial communities revealed by the ego networks of all the vertices.}
   \label{fig:first_neighbourhood}
\end{figure}

Structurally, communities with significant overlaps are hidden
under dense and messy edges, unlike the cases of disjoint and
slightly overlapping communities.  Identifying such communities is
highly non-trivial from a global or top-down viewpoint. Starting locally from a vertex, however, it is easier to
identify which groups a vertex belongs to in the subnetwork
consisting of the vertex itself and its neighbors, i.e. the ego
network of a vertex, as illustrated in
Fig.~\ref{fig:first_neighbourhood} using data from an online
social network.
Several community detection algorithms that take advantage of such local community structure in ego networks have been developed in recent years. Coscia~\textit{et al.} proposed a method DEMON  \cite{Coscia:2012ip, Coscia:2014cw} that let each vertex vote for the communities it sees in its ego network democratically and then merges similar communities repeatedly if the overlap between two communities is above a threshold. The method EgoClustering by Rees \textit{et al.}~\cite{Rees:2012jb}  follows a very similar idea with a different merger criterion. Later on Soundarajan and Hopcroft proposed the Node Perception algorithm~\cite{Soundarajan:2015gq}, which is a general algorithm template consisting of three steps: firstly detect subcommunities in the neighborhood of each vertex; then create a new network of which each vertex represents a subcommunity and the edges represent some relationship between subcommunities; and finally detect communities in the network of subcommunity. 
Since the merger of similar subcommunities can be reinterpreted as the second and the third step, DEMON and EgoClustering can be regarded as implementations and special cases of the template. This intuitive local-first approach is easy to conceive. The key to success lies in how to implement it, especially for the latter two steps, i.e. finding a proper way of merging similar communities. There are two key issues to be addressed: merger order and noise control. In DEMON, a community $A$ is immediately merged with another community $B$ if the overlap between them is above a threshold, without considering whether there exists a better candidate $C$ to be merged with $A$. This would lead to less accurate results, and we think always merging with the best candidate is preferable. Also, the detected communities in ego networks may contain noise, i.e. vertices that are misclassfied into these communities. The noise accumulates quickly as the merger proceeds and the similarity measure, i.e. the overlap between two merged communities used in DEMON and EgoClustering, becomes less and less accurate. It eventually leads to poor accuracy, as we will show in benchmarking in Sec.~\ref{sec:benchmarks}. Taking these aspects into account, we develop a Partial Community Merger Algorithm (PCMA) based on the same local-first approach for detecting communities with significant overlaps as well as slightly overlapping and disjoint ones. We propose a sophisticated yet efficient similarity measure between merged communities that can trace the merger history and suppress noise accumulated during the process. The mathematical properties of the measure allow us to design a linear complexity merger algorithm that always merges the most similar pair of communities. The variables used in the measure can be utilized to establish a set of thresholds for further reducing noise after merger. Besides the efficiency and accuracy ensured by the proposed similarity measure, our method is also equipped with many features that are essential to dealing with large-scale real networks. Like other algorithms based on local approaches, the method does not require an input of the total number of communities to be detected.
Most importantly, the assumption of disjoint or slightly overlapping
communities is abandoned and the method is designed to handle the
possibility of multiple memberships of a vertex and detect
significantly overlapping communities.  Our method also allows
vertices to be isolated, i.e., they do not belong to any
communities. Community detection
algorithms must be able to distinguish real communities from
pseudo communities~\cite{Bianconi:2009il,Lancichinetti:2010ck}. We
are well aware of the issue and our method sifts out real
communities by applying proper thresholds.  All these advantages
make the method uniquely capable of detecting communities with
significant overlaps efficiently in large-scale real networks with
hundreds of millions of vertices.

The plan of the paper is as follows.  In
Sec.~\ref{sec:the_method}, we introduce the details of the
algorithm, including similarity measure, merger of similar
communities, thresholds, and applicability.  In
Sec.~\ref{sec:benchmarks}, the method is tested against two
benchmarks and its performance in accuracy and efficiency is
compared with four other recently proposed algorithms. In
Sec.~\ref{sec:empirical_analysis}, we apply the method to two huge social networks and compare the communities detected with the metadata groups. We show that significantly overlapping communities are common in social networks. Results are summarized
in Sec.~\ref{sec:summary}.

\section{Partial Community Merger Algorithm}
\label{sec:the_method}
\subsection{General Principle}
We aim to detect communities in a network in which vertices could
span from being isolated to belonging to multiple communities.
This renders many top-down algorithms invalid.  We first give a
physical picture of our algorithm.  Consider a community in which
every member is connected to a certain fraction of the other
members. At the local level of the members, they only known their
own neighbors and have no knowledge of the complete community.
They are given the task of compiling a roster of the community and
identify who the core members are.  To complete the task, each
member shares its local information consisting of a name list
including itself and all its neighbors.  A complete roster can in
principle be derived by merging these individual name lists
skillfully.  Those who appear frequently on the lists are the core
members, while those with less occurrence are on the periphery of
the community.  This merger process is the core idea of our method
of detecting communities.

Practically, we start with exploring the ego network of a vertex,
i.e. the subnetwork consisting of the vertex itself and its
neighbors, and identifying the communities hidden in it.  This is
illustrated in Fig.~\ref{fig:first_neighbourhood} for a vertex
(the central one) in an online social network.  This local view
lets us see the communities clearly.  Since a vertex may not know
all the members in each of its communities, the identified communities in the ego network of a vertex are
incomplete. We refer to them as the \textit{partial communities}
from the viewpoint of the vertex. This process can be carried out
for every vertex. Although each member only helps reveal part of
the whole picture, the idea is that aggregating local information
should reveal the \textit{complete communities}, i.e., every
community with all its members.  With the partial communities
revealed by different vertices, the next step is to determine which ones
are actually different parts of the same community. It is a technically difficult task as the number of partial communities may even be greater than the number of vertices in the whole network and vertices may be
misclassified into partial communities that they actually do not
belong to. Merger of corresponding partial communities in principle reconstructs the complete community. It, however, also causes the noise in partial communities to be accumulated. A cleaning process or post-processing scheme is then
invoked to eliminate the misclassified vertices and sift out the
real and complete communities.

Our method thus consists of three steps:
\begin{enumerate}
\item Find the partial communities in the ego network of each of
the vertices.

\item Merge partial communities that are parts of the same
community to reconstruct complete communities.

\item Clean the merged communities to sift out real communities.
\end{enumerate}
For easy reference, we call the method Partial Community Merger
Algorithm or PCMA in short. For Step 1, many existing
algorithms are available and we use the one proposed by Ball
\textit{et al.} \cite{Ball:2011wz}, with details given in
Appendix~\ref{sec:appendix_1}.  The new elements are Step 2 and
Step 3.  Below, we introduce our implementation of Steps 2 and 3
in detail.

\subsection{PCMA Step 2: Merger}
\label{subsec:merger}
The merger process aims to reassemble partial communities that are part of the same complete community. As discussed in Sec.~\ref{sec:introduction}, care must be taken over the merger order and noise control to make the process effective. We propose an efficient hierarchical clustering based algorithm that always merges the most similar pair of communities and it can handle noise properly.

The key here is to find a suitable similarity measure
between communities, each of which is represented by a set of vertices. The Jaccard index is a commonly used similarity measure. It is defined to be
the size of the intersection divided by the size of the union of
two sets.  A drawback of the index is that the members are assumed
equal.  A merged community, however, contains core members,
peripheral members, and even misclassified ones.  They should be
treated differently.  Here we propose a novel similarity measure
that incorporates the different importance of members and hence reduce significantly the interference of the misclassified vertices on the measure.

For a Vertex $i$ in Community $C$, let $S_{i,C}$ be a score that
represents its importance in $C$.  Without loss of generality, we
define $S_{i,C}=0$ if $i \notin C$.  Let $l_{C}$ be the number of
partial communities that have merged to form Community $C$. Before
the merger, the partial communities identified by Step 1 all have
$l=1$ and all members carry an initial score of $1$. When two
communities $A$ and $B$ merge into one, e.g. $C=A\cup B$, the
quantities $S_{i,C}$ and $l_{C}$ are given by
\begin{eqnarray}
    S_{i,C} &= S_{i,A} + S_{i,B} \\
    l_C &= l_A + l_B
  \label{eq:merge_score_l}
\end{eqnarray}
Physically, $S_{i,C}$ traces the number of occurrences of Vertex
$i$ in the $l_C$ partial communities that have merged to form
Community $C$.  Vertices with a high value of $S/l$ are regarded
as core members, and those with a small $S$, say less than $3$,
are very likely vertices that are misclassified.

Consider two communities $A$ and $B$.  We define an asymmetric
measure $f(A,B)$ to take into account the different importance of
members as
\begin{equation}
  f(A,B) = \sum\limits_{i}{\frac{S_{i,B}}{l_B} \cdot \frac{S_{i,A}}{w_A}} \;,
  \label{eq:f_asym}
\end{equation}
where $w_A=\sum\limits_{i}{S_{i,A}}$. The term $S_{i,B}/l_{B}$
represents a normalized importance of Vertex $i$ in $B$, ranging from $0$ to $1$, and
$S_{i,A}/w_{A}$ is a weighting factor of Vertex $i$ in $A$. So $f(A,B)$ measures the weighted average importance of $A$'s members in $B$. The
product $S_{i,B} \cdot S_{i,A}$ ensures that $f$ will not be
affected much by the misclassified vertices, i.e. those with small
values of $S$. A large value of $f(A,B)$ indicates that the core
members of $A$ are also core members of $B$, but \textit{not} vice
versa as $f(A,B) \neq f(B,A)$ in general.  This measure has the
following properties:
\begin{equation}
  f(A, B\cup C) = \frac{l_B}{l_B+l_C} f(A,B) + \frac{l_C}{l_B+l_C} f(A,C)
  \label{eq:f_3}
\end{equation}
\begin{equation}
  f(B\cup C, A) = \frac{w_B}{w_B+w_C} f(B,A) + \frac{w_C}{w_B+w_C}
  f(C,A)\;,
  \label{eq:f_4}
\end{equation}
which can be readily shown.  Let $\{A\}$ ($\{B\}$) denote the set
of partial communities that form the Community $A$ ($B$).  It
follows from Eqs.~(\ref{eq:f_3}) and (\ref{eq:f_4}) that
\begin{equation}
    f(A,B) = f(\cup_{x\in \{A\}}{x}, \cup_{y\in \{B\}}{y})
    = \sum_{x\in \{A\} \atop y\in \{B\}}\nolimits{\frac{w_x}{w_A l_B}
    f(x,y)}\;.
  \label{eq:f_5}
\end{equation}
Recall that $x$ and $y$ are partial communities and thus $f(x,y)$
is the portion of members of $x$ who are also members of $y$, i.e.
\begin{equation}
  f(x,y) = \frac{\mid x \cap y \mid}{\mid x \mid} \;.
\end{equation}
Equation~(\ref{eq:f_5}) indicates that $f(A,B)$ is actually a
weighted average of the overlap portion $f(x,y)$ over all
combinations of partial communities forming $A$ and $B$, i.e. with
$\{(x,y), x\in\{A\}, y\in\{B\}\}$.

The merger of two communities $A$ and $B$ is different from either
$A$ absorbing $B$ or $B$ absorbing $A$.  Thus, a symmetric analogy
of $f(A,B)$ is preferred for deciding a merger.  To motivate the
construction of such a parameter, we introduce a measure $g(C)$ of
a Community $C$ in a way similar to Eq.~(\ref{eq:f_5}) that
compares members in the partial communities forming $C$:
\begin{equation}
g(C) = \left\{ 
  \begin{array}{l l}
    1 & \quad \textrm{if} \ l_C=1\\
    \sum_{x,y \in \{C\} \atop x \neq y}
    \frac{w_x f(x,y)}{w_C(l_C-1)} \ & \quad \textrm{if} \ l_C > 1
  \end{array} \right.
\end{equation}
It gives the average portion of overlap between partial
communities in a merged community.  Its physical meaning can be
seen by considering the special case that $C$ is an ER random
network $G(n,p)$ with members randomly connected with a
probability $p$.  A partial community now consists of a vertex and
all its neighbors.  The expected portion of overlap between two
partial communities $f(x,y)$ is roughly $p$, giving $g(C) \approx
p$.  This indicates that $g(C)$ is approximately a measure of the
fraction of other members that a member is connected to.  A larger
$g(C)$ implies denser internal edges in the community and thus
members are connected tightly to each other.  It can be used as an
indicator on whether a merged community is a real community or
just a wrongly merged set of vertices.

For the case $C=A\cup B$, $g(C)$ satisfies
\begin{eqnarray}
  g(C) & = \frac{1}{w_C(l_C-1)} \left\{ w_A (l_A-1)g(A) + w_B(l_B-1)g(B) \right. \nonumber \\
          & \quad \left.  + (w_A l_B + w_B l_A) f_s(A,B)  \right\}
  \, ,
\label{eq:g_cab}
\end{eqnarray}
where
\begin{equation}
f_s(A,B) = \frac{w_A l_B f(A,B) + w_B l_A f(B,A)}{w_A l_B + w_B l_A}
              = \frac{2 \sum_i S_{i,A} \cdot S_{i,B}}{w_A l_B + w_B l_A} \;.
\label{eq:fs}
\end{equation}
There are three terms in Eq.~(\ref{eq:g_cab}) for $g(C)$.  The
first and second terms give the overlap portions within $A$ and
within $B$, respectively.

The third term in Eq.~(\ref{eq:g_cab}) measures the overlap
between $A$ and $B$.  It is important to note that a {\em
symmetric measure} $f_{s}(A,B)$, as defined in Eq.~(\ref{eq:fs}),
emerges. It is a weighted average of the asymmetric measures
$f(A,B)$ and $f(B,A)$ and yet itself satisfies $f_{s}(A,B) =
f_{s}(B,A)$.  It follows from Eq.~(\ref{eq:fs}) that $f_s(A,B \cup
C)$ is given by a weighted average of $f_s(A,B)$ and $f_s(A,C)$,
and thus
\begin{equation}
f_s(A, B \cup C) \leqslant \max \left\{ f_s(A,B), f_s(A,C)
\right\} \;. \label{eq:fs_u}
\end{equation}
We are thus led to apply $f_{s}$ as a symmetric similarity measure
between two communities that accounts for the different importance
of the members. We remark that $f_{s}$ is not the only option but it is a good one because of its many advantages: $f_{s}$ has a clear physical meaning that measures the average overlap portion of partial communities between $A$ and $B$; it respects the different importance of members of a community; it is linear and easy to calculate; and its mathematical property expressed in Eq.(~\ref{eq:fs_u}) allows us to optimise the merger process to be of linear complexity as we will discuss shortly.

Based on the idea of agglomerative hierarchical clustering, the
merger process using $f_s$ as the similarity measure can be
implemented as follows. Given a set $\mathscr{C}$ of communities
to be merged, a straightforward way is to:
\begin{enumerate}
\item Calculate $f_s$ for each pair of communities in
$\mathscr{C}$ and maintain a priority queue of $f_s$ in descending
order.

\item Merge the pair with the largest $f_s$ and update the
priority queue.

\item Repeat 2 until the largest $f_s$ in the priority queue falls
below a threshold $t_{f_s}$.
\end{enumerate}
The time complexity of this algorithm is $O(n^2 \log n)$, where
$n$ is the number of communities in $\mathscr{C}$. The space
complexity is $O(n^2)$ as we need to maintain the priority queue
of $f_s$.  For detecting communities in large-scale networks, a
more efficient algorithm is desirable.  In what follows, we
propose two optimizations to reduce both the time and space
complexity to $O(n)$.

We define the \textit{best merger candidate} of a community $A$ as
\begin{equation}
 \mathrm{bmc}(A) = \mathop{\arg \max}\limits_{X \in \mathscr{C} / A } f_s(A, X) \;.
\label{eq:bmc}
\end{equation}
We argue that the algorithm above is equivalent to:
\begin{algorithmic}[1]
\State \textbf{given} a set of communities $\mathscr{C}$
\Repeat
   \State choose a community $A$ from $\mathscr{C}$
   \State $B \gets \mathrm{bmc}(A)$
   \While{$\mathrm{bmc}(B) \neq A$}
        \State $A \gets B$
        \State $B \gets \mathrm{bmc}(A)$
   \EndWhile
   \If{$f_s(A,B) > t_{f_s}$}
       \State merge $A$ and $B$
       \State remove $A$ and $B$, add $A \cup B$ to $\mathscr{C}$
   \EndIf
\Until{ no communities can be merged anymore}
\State \Return $\mathscr{C}$
\end{algorithmic}
The algorithm makes use of the property of $f_{s}$ given in
Eq.~(\ref{eq:fs_u}).  If $A$ and $B$ are the best merger
candidates of each other, there does not exist a community $C$
that gives $f_s(A,C) > f_s(A,B)$, where $C$ can be any combination
of communities in $\mathscr{C}/A$.  Therefore, even if $f_s(A, B)$
is not at the top of the $f_s$ priority queue, the merger of $A$
and $B$ can be moved forward since other mergers higher on the
priority queue that would take part will not affect the merger of
$A$ and $B$.  An advantage is that merges are not required to
proceed in order in the algorithm, and thus there is no need to
maintain the $f_s$ priority queue. The space complexity is reduced
from $O(n^2)$ to $O(n)$.

The search on $\mathrm{bmc}(A)$ is formally within the set
$\mathscr{C}/A$.  Practically, the search area can be reduced
significantly, as most of the communities in $\mathscr{C}/A$ do
not even share a single member with $A$ in sparse networks. A good
approximation is to limit the search to the partial communities
from the viewpoints of $A$'s members and the merged communities
containing these partial communities.  As such, the time
complexity of calculating $\mathrm{bmc}(A)$ does not scale
with $n$, providing that the community size and the number of
partial communities per vertex are independent of the network
size.  The number of iterations of finding a pair of communities
to merge should also be independent of $n$.  The repeated loop
requires a time complexity of $O(n)$. Since $n$ usually scales
linearly with the network size, it can also be regarded as the
network size.  We thus argue that the time complexity of our
optimized merger algorithm is approximately $O(n)$.  This is
verified numerically in Sec.~\ref{sec:benchmarks}.

The choice of the value of $t_{f_s}$ depends on the networks being analyzed, and one's subjective view of how dense the internal edges of a community should be. There is no standard answer. For example, setting $t_{f_s}=0.1$ means we aim to detect communities of which the members know more than $10\%$ of the other members on average. Using a higher value means PCMA will focus on communities with denser internal edges or the cores inside them, and those with a lower density are not acknowledged as communities. Lowering $t_{f_s}$ would bring in more peripheral vertices as members and would cause further merger of similar communities. Tuning $t_{f_s}$ from high to low may extract the core-periphery and possible hierarchical structure of communities. It is another important topic that is beyond the scope of the present work, we leave it to future work. We think $t_{f_s}=0.1$  is a good choice for social communities and it is harsh enough for those with a large size. However, it can easily be satisfied by small ones. For example, two partial communities of size $10$ only need one common member to satisfy the threshold. Since partial communities detected in PCMA Step 1 may contain a lot of misclassified members, such loose threshold for small partial communities may result in a lot of mergers of partial communities that are actually not part of the same complete community. We suppress such unwanted mergers of small partial communities by forcing the similarity $f_s\left(A,B\right)=0$ if $\sum_i S_{i,A} \cdot S_{i,B} / \max \{l_A, l_B\} < t_{f_0}$. It means two partial communities can merge only if the number of their common members is no less than $t_{f_0}$. A better solution is to use a dynamic $t_{f_s}$ that depends on the sizes of communities and we leave it to future work.

\subsection{PCMA Step 3: Post-processing}
\label{subsec:post-processing}

After merging communities, a cleaning process is needed to handle
two types of noise.  First, we need to identify which merged
communities are real communities and which are simply merged sets
of vertices by coincidence.  The latter usually contain only a
small number of partial communities because they are merged by
accidents.  The more partial communities a merged community
contains, the more likely it is a real community. Thus, the
parameter $l$ of a community can be used as a measure of whether a
detected community is trustful.  A way to sift out real
communities is to set a threshold $t_l$ and require all real
communities to have $l \geqslant t_l$.  The threshold can be set in
many ways, e.g. setting $t_l$ based on each community's size, as a
larger community usually caries a larger $l$.  Second, we need to
identify and eliminate vertices that are misclassified into
partial communities in Step 1.  Recall that $S_{i,C}$ is the
number of occurrences of Vertex $i$ in $l_C$ partial communities
that formed the Community $C$. Excluding Vertex $i$'s own partial communities (if any), the remaining score $S'_{i,C}$ can roughly be interpreted as the number of other members that Vertex $i$ is connected to in Community $C$. Usually $S'_{i,C}$ = $S_{i,C}-1$. If the probability of Vertex $i$ being misclassified into a partial community is $10\%$, then being misclassified into $S'_{i,C}$ partial communities of $C$ is $0.1^{S'_{i,C}}$. Therefore, the probability of Vertex $i$ being a false member drops sharply with increasing $S'_{i,C}$.  Thus, a threshold $t_{S'}$ can be set to eliminate vertices with $S' < t_{S'}$.
Normally $t_{S'}=4$ is sufficiently stringent and it should not be
less than $2$.  There remain vertices with $S' \geqslant t_{S'}$ but $S'/l
\approx 0$.  They may still not be members since they know too few
other members.  The ratio $S'/l$ gives an estimate on the fraction
of the other members that a member is connected to.  Another
criterion $S'/l > t_{S'/l}$ becomes useful, with $t_{S'/l}$ being a
threshold that requires each member be connected to at least
$t_{S'/l} \times 100\% $ of the other members.  This criterion
echoes the concept of $f$-core discussed in
Sec~\ref{sec:introduction}.  The threshold $t_{S'/l}$ can either be
set uniformly for all communities or individually for each
community based on the value of $g$, which reflects the average
portion that a member is connected to the other members.  Since
different kinds of networks may have different community
structures, the choice of $t_{S'/l}$ depends on the nature of
communities of a specific network under study.

\subsection{Applicability}
\label{subsec:applicability}

PCMA works under two conditions: Existence of partial communities
(Step 1) and adequate overlap between partial communities for
mergers (Step 2).  Usually, the second condition is satisfied
automatically when the network under study meets the first
condition.  For the first condition, the existence of partial
communities from the viewpoint of a vertex requires that there are
sufficient number of neighbors and a high density of edges among
the neighbors, i.e., a high local clustering coefficient of the
vertex.

We expect a community detected by PCMA to have the following
properties:
\begin{enumerate}
\item Two members with common neighbors are highly likely
neighbors of each other.  As a consequence, the community has a
relatively high value of clustering coefficient.

\item The shortest distances between most pairs of members are
generally short and not longer than 3.  Thus, most members are
connected to each other either directly or via one/two
intermediate member(s).

\item Each member is connected to at least a certain fraction of
the other members.
\end{enumerate}
From another perspective, these properties can be taken as a broad
descriptive definition of community, and are well suited for
describing communities with significant overlaps.  PCMA is
designed to detect this kind of communities, which are important
in large-scale systems with vertices typically having multiple
memberships.

\section{Benchmarking}
\label{sec:benchmarks}

We tested PCMA using two benchmark models to illustrate its
performance and applicability.   Results are compared with DEMON~\cite{Coscia:2012ip}, which shares the same general idea as that of PCMA, and also three other
fast and accurate overlapping community detection algorithms that
are among the best~\cite{Xie:2013ku}: namely
OSLOM~\cite{Lancichinetti:2011gn}, SLPA~\cite{Xie:2011hc}, and BIGCLAM~\cite{Yang:2013ko}.

First, a simple benchmark model in the spirit of the planted
$\ell$-partition model~\cite{Condon:2001jj} is used.  The network is
generated as follows:
\begin{enumerate}
\item Generate an ER random network of $n$ vertices with a mean
degree $\left<k\right>$ that serves as background noise.

\item Randomly sample $s$ vertices as a community, with $s$
satisfying the Poisson distribution with an expected value of
$\left<s\right>$.  Connect each pair of members with a probability
$p$.

\item Repeat the step to generate $n\cdot \left<c\right>$
communities. Here, $\left<c\right>$ is the expected number of
communities that a vertex belongs to.
\end{enumerate}
This model is flexible in that the total number, size, and
intra-community edge density, as well as the background noise
level can be directly controlled. In addition, vertices can belong
to an arbitrary number of communities, including zero. There is no guarantee
that there are more edges within a community than edges going out.
These features make many existing community detection algorithms
invalid.  They reflect the challenges posed by real social
networks, in which a person often simultaneously belongs to many
groups, on community detection. PCMA is designed to solve the
problem.

\begin{figure}[htbp]
   \centering
   \epsfig{file=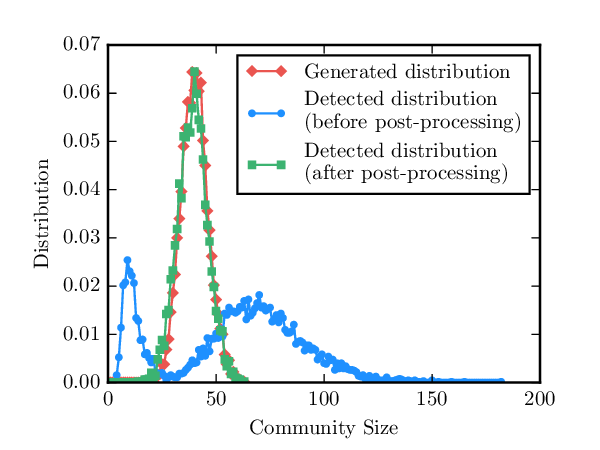, width=0.7\linewidth}
   \caption{(Color online) Community size distribution as
   generated by the benchmark model (diamonds) described in the text.
   The corresponding results as detected by our method before (circles) and after
   (squares) applying the post-processing scheme are shown for
   comparison.  The thresholds are given in Table~\ref{tab:threshold_pcma_2_3} in Appendix~\ref{sec:appendix_2}.}
   \label{fig:bm_simple}
\end{figure}

Consider a network with $n=10^5$, $\left<k\right>=20$, $p=0.3$,
$\left<s\right>=40$, and $\left<c\right>=2$ generated as
described.  The threshold $t_{f_s}=0.1$ is chosen for the merger
process.  Fig.~\ref{fig:bm_simple} shows the actual community
size distributions generated by the model (diamonds). The results
as detected by PCMA before (circles) and after post-processing
(squares) are shown for comparison.  As discussed in
Sec.~\ref{subsec:post-processing}, the merged communities without
post-processing are noisy.   The results show a peak at a small
community size due to many coincidentally merged sets of vertices,
which are targeted for removal in post-processing.  The results
also show a bump in the distribution at large community sizes.
Although these are real communities, the many misclassified
vertices make their sizes bigger than their actual sizes.  Hence,
results before post-processing could be misleading. By setting the
thresholds properly, as given in Table~\ref{tab:threshold_pcma_2_3} in Appendix~\ref{sec:appendix_2}, the distribution after post-processing
is in good agreement with the actual community size distribution.

To qualify the accuracy of PCMA, we adopt the widely used
Normalized Mutual Information (NMI)~\cite{Danon:2005gr} as
extended by Lancichinetti {\em et al.}~\cite{Lancichinetti:2009dy}
to compare overlapping communities.
As a recent study showed there may be cases in which NMI is biased~\cite{Zhang:2015bo}, we use two additional measures, namely the Omega Index~\cite{Murray:2012tp} and $F_1$ Score, to make the evaluation more comprehensive. The $F_1$ Score in the present work is defined as
 \begin{equation}
   F_1(\mathscr{D}, \mathscr{G}) = \frac{2\cdot Precision(\mathscr{D}, \mathscr{G}) \cdot Recall(\mathscr{D}, \mathscr{G})}{Precision(\mathscr{D}, \mathscr{G}) + Recall(\mathscr{D}, \mathscr{G})} \;,
  \label{eq:f1}
\end{equation}
 \begin{equation}
   Precision(\mathscr{D}, \mathscr{G}) = \frac{1}{|\mathscr{D}|} \sum_{D \in \mathscr{D}} \max_{G \in \mathscr{G}} \frac{|D \cap G|}{|D \cup G|} \;,
\end{equation}
 \begin{equation}
   Recall(\mathscr{D}, \mathscr{G}) = Precision(\mathscr{G}, \mathscr{D}) \;,
\end{equation}
where $\mathscr{D}$ and $\mathscr{G}$ are the detected communities and the ground-truth, respectively.

\begin{sidewaysfigure}
   \centering
   \epsfig{file=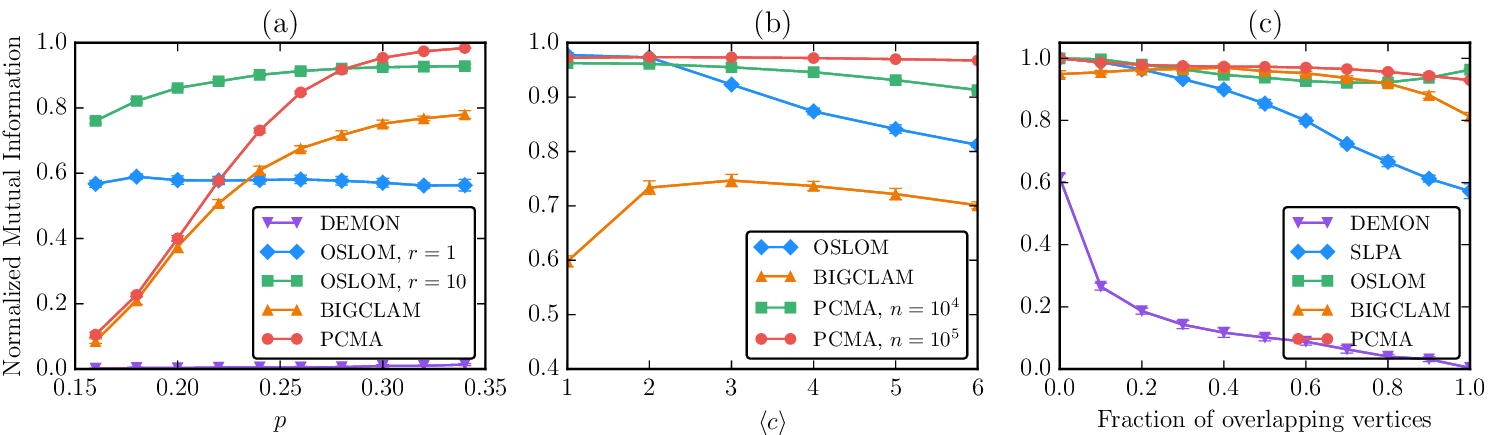, width=\linewidth}
   \vspace{0.5\baselineskip}
   \caption{(Color online) Performance comparisons of DEMON, SLPA, OSLOM, BIGCLAM, and PCMA in a simple benchmark model and the LFR benchmark model.  Unless stated otherwise, parameters of our simple benchmark model
in (a) and (b) are: $n=10^4$, $\left<k\right>=20$, $p=0.3$,
$\left<s\right>=40$, $\left<c\right>=3$.  Parameters of the LFR
benchmark model in (c) are: $n=10^4$, $\left<k\right>=40$,
$k_{max}=100$, $\mu=0.3$, $\tau_1=2$, $\tau_2=1$, $c_{min}=20$,
$c_{max}=100$, each overlapping vertex has two communities. The thresholds of PCMA are given in Table~\ref{tab:threshold_pcma_2_3} in Appendix~\ref{sec:appendix_2}. For (b), the number of partial communities set to be found in PCMA Step 1 is $K=10$. For DEMON, the default value of the parameter $\epsilon=0.25$ is used. The number of iterations of OSLOM is
set to $r=10$.  For SLPA, the program applies different thresholds
ranging from $0.01$ to $0.5$ by default and we select the best
result. BIGCLAM is informed of the actual number of communities generated by the benchmark models.
Each data point is an average of $10$ realizations.  If
not shown, the error bar is smaller than the size of the symbol.}
   \label{fig:nmi}
\end{sidewaysfigure}

\renewcommand{\thefigure}{\arabic{figure} (Cont.)}
\addtocounter{figure}{-1}
\begin{sidewaysfigure}
   \centering
   \epsfig{file=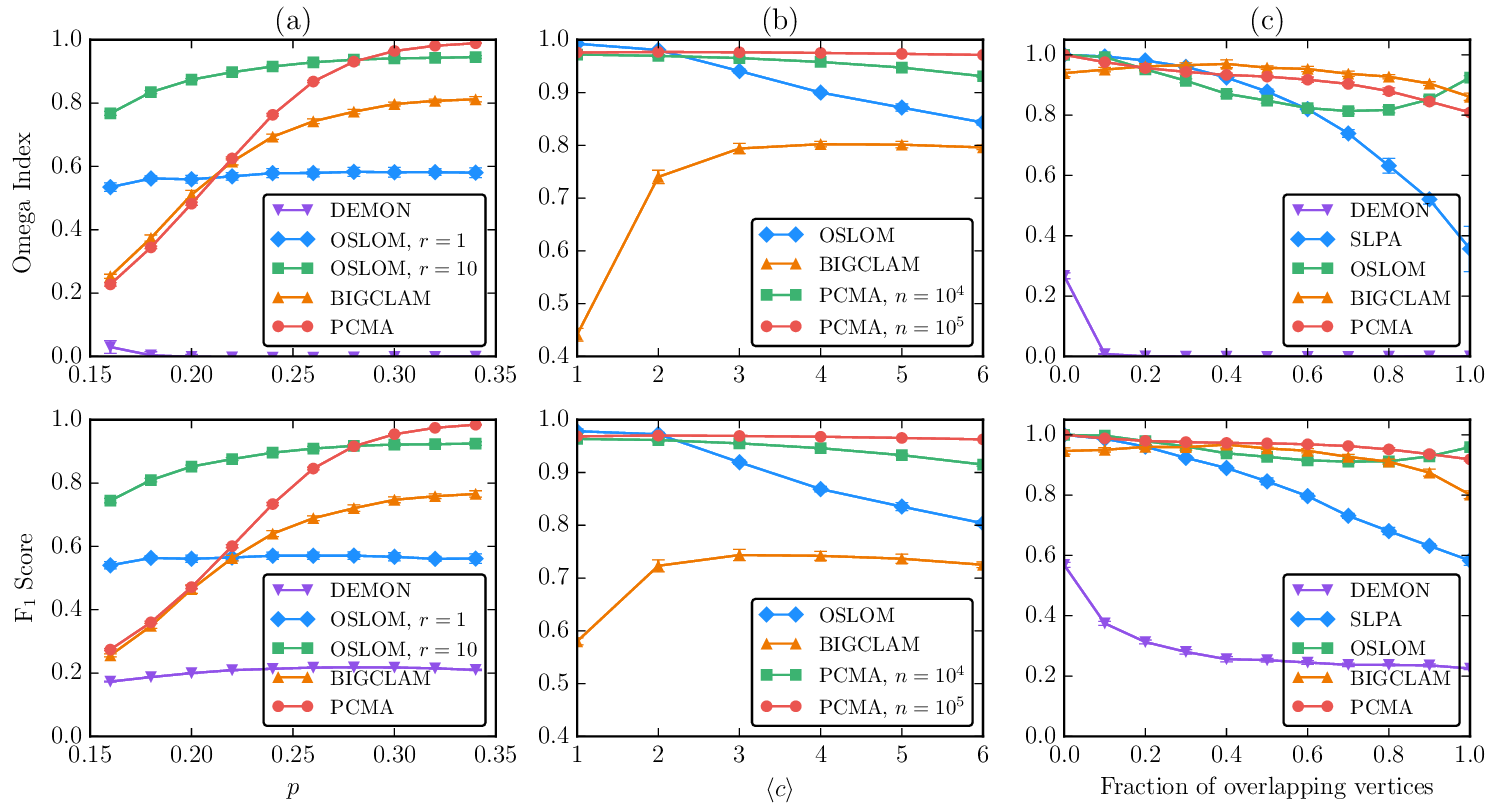, width=\linewidth}
   \vspace{0.5\baselineskip}
   \caption{(Color online) Performance comparisons of DEMON, SLPA, OSLOM, BIGCLAM, and PCMA in a simple benchmark model and the LFR benchmark model (continued).  All parameters are the same as those stated in the caption of Fig.~\ref{fig:nmi}.}
   \label{fig:omega_f1}
\end{sidewaysfigure}
\renewcommand{\thefigure}{\arabic{figure}}

Fig.~\ref{fig:nmi}(a) compares the performance of
DEMON, OSLOM, BIGCLAM, and PCMA on synthetic networks
with different intra-community edge densities.  Results of SLPA
are not shown because the method cannot detect isolated vertices.

DEMON failed to detect communities generated by this benchmark as values of the NMI and Omega Index are almost zero. As discussed in Sec.~\ref{sec:introduction}, a main reason is that the misclassified vertices invalidate the similarity measure in DEMON which is based on the overlap between two communities. Fig.~\ref{fig:c_before_step_3} shows an example of a merged community containing two times more misclassified vertices than true members. It indicates clearly that a similarity measure capable of tracing the different importance of vertices in a merged community, such as the one used in PCMA, is essential for the effectiveness of the merger.

The accuracy of OSLOM depends strongly on the number of
iterations.  We use the default value $r=10$ suggested in
Ref.~\cite{Lancichinetti:2011gn}, unless specified otherwise.
BIGCLAM and PCMA are shown to be sensitive to the intra-community edge density $p$. For
PCMA, it is 
because $p$ affects the existence of partial communities, which is
a criterion for the applicability of PCMA (see
Sec.~\ref{subsec:applicability}).  In the benchmark model, the
probability $p$ promotes partial communities.  PCMA works well
when
\begin{equation}
  \left<k_{nn}\right> = \left[(\left<s\right>-1)p-1 \right]p \geqslant
  2 \;,
  \label{eq:pc_c}
\end{equation}
where $(\left<s\right>-1)p$ is the expected number of neighbors of
a member and $\left<k_{nn}\right>$ is the expected number of edges
of a neighbor connecting to other neighbors of the member. The
neighbors start to be strongly connected, i.e. partial communities
emerge, when $\left<k_{nn}\right> \geqslant 2$. From
Fig.~\ref{fig:nmi}(a), PCMA performs better than OSLOM for $p >
0.28$, corresponding to $\left<k_{nn}\right> > 2.78$ for the
case of $\left<s\right>=40$.

Fig.~\ref{fig:nmi}(b) shows the dependence of the performance on
$\left<c\right>$, which controls the expected number of
communities that a vertex belongs to.  A larger $\left<c\right>$
corresponds to more edges connecting the communities and thus a
denser and more complex network. The performance of DEMON is far worse than the other methods (similar to what is shown in Fig.~\ref{fig:nmi}(a)) and the result is not shown. For a system with $n=10^{4}$
members, the accuracy of OSLOM falls rapidly with increasing
$\left<c\right>$.  For PCMA, the accuracy remains high throughout,
with a slight drop due to the finite size of the network instead
of $\left<c\right>$.  This is verified by the performance of PCMA
in a bigger system of $n=10^{5}$ (circles in
Fig.~\ref{fig:nmi}(b)).  Recall that many existing algorithms
become invalid in problems that a vertex may belong to many
communities, but PCMA handles them well.

We also tested PCMA with the widely used LFR benchmark
model~\cite{Lancichinetti:2008ge}.
The results are shown in Fig.~\ref{fig:nmi}(c). In the LFR
model, a vertex has a degree chosen from a distribution that
follows a power-law of exponent $\tau_{1}$ in a range of degrees
$k_{min} \leqslant k \leqslant k_{max}$ corresponding to a mean degree
$\langle k \rangle$.   A tunable fraction of vertices are chosen
to belong to more than one communities.  They are the overlapping
vertices.  The remaining vertices have only one community.  For a
vertex of degree $k$, a parameter $\mu$ sets the fraction of the
edges to be connected to vertices outside the community(ies) that
the vertex belongs to. The remaining fraction $(1 - \mu)$ of edges
are evenly divided among the communities, if the vertex is chosen
to have multiple communities. As such, the community sizes also
follow a power-law with an exponent $\tau_{2}$ within a range of
communities sizes between $c_{min}$ and $c_{min}$.  The
combinations of parameters in the LFR model give a class of
tunable structures for the resulting networks.
The parameters used in benchmarking are based on the empirical networks we studied.  We use the mean degree $\left<k\right>=40$ of the two real social networks shown in Table~\ref{tab:social_networks_analyzed}. The selection of the mixing parameter $\mu$ and the range of community size is close to what we observed in the two networks.
Fig.~\ref{fig:nmi}(c) shows OSLOM, SPLA, and PCMA work very well when
there are very few overlapping vertices.  When communities overlap
more, PCMA achieved higher values of the NMI and $F_1$ Score than the other four methods over a wide
range of the fraction of overlapping vertices, except for the last
data point in Fig.~\ref{fig:nmi}(c) in comparison with OSLOM.  In
the LFR model, the degree assignment does not distinguish
single-community vertices from multi-community vertices.  As a
result, a vertex belonging to multiple communities has fewer edges
connecting to each of its communities.  In PCMA, however, members
are expected to be connected to at least a certain fraction of the
other members before establishing their membership. This leads to
the gradual drop in PCMA's NMI with increasing number of
overlapping vertices, which are members according to the benchmark
model but may not be acknowledged by PCMA.  We remark that it is
actually not a problem of accuracy, but more about what a
community should be.

\begin{figure}[htbp]
   \centering
   \epsfig{file=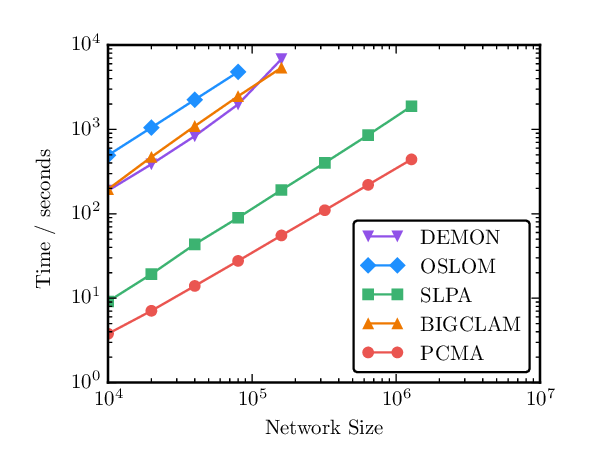, width=0.7\linewidth}
   \caption{(Color online) Comparison on time complexity.
   Tests are conducted on LFR benchmark model. The fraction of overlapping
   vertices is set to $50\%$. Other parameters are the same
   as those in Fig.~\ref{fig:nmi}. Each data point is an average
   over $10$ realizations. Error bars are smaller than the size of the symbols.}
   \label{fig:time}
\end{figure}

We also studied the time complexity of the methods
numerically based on the LFR benchmark
model~\cite{Lancichinetti:2008ge}.  Calculations were performed on
a workstation with Intel Xeon E5-2609 @ 2.40GHz (4 cores / 8
threads).  The programs were allowed to use all threads if they
were parallelized. Fig.~\ref{fig:time} shows how the execution
time scales with the network size.  SLPA and PCMA are very efficient, while OSLOM and BIGCLAM are a few hundred times slower. DEMON is written in Python, which by nature is much slower than programs written in C/C++. In the
log-log plot in Fig.~\ref{fig:time}, the slopes for SLPA, OSLOM, BIGCLAM,
and PCMA are $1.09$, $1.09$, $1.20$, and $0.99$, respectively.  It is,
therefore, numerically verified that the time complexity of PCMA
is $O(n)$.

In summary, the benchmark tests showed that PCMA is an efficient
algorithm specifically suitable for detecting communities in
networks in which the vertices may belong to multiple communities.

\section{Empirical Analysis}
\label{sec:empirical_analysis}

Having established the efficiency and accuracy of PCMA, we tested it against two huge online social networks, Friendster~\cite{Yang:2013jw} and Sina Weibo. The basic information is listed in Table~\ref{tab:social_networks_analyzed}. Friendster is one of the earliest social network service websites, allowing people to maintain contacts and interact with each other. The network data of Friendster is downloaded from SNAP Datasets~\cite{snapnets}. It comes with \textit{metadata groups}~\cite{Hric:2014jt}, or the so called \textit{ground-truth} communities~\cite{Yang:2013jw, Yang:2014fc}, providing us with a reference to evaluate the accuracy of detection algorithms. Sina Weibo is a directed network like Twitter for fast information spreading and it also has characteristics of Facebook for interactions with friends. We focused on the embedded undirected friendship network with only reciprocal edges. The network we sampled from the Internet contains about $80$ million vertices and $1.0$ billion reciprocal edges, with only $1.2\%$ of the edges being connected to vertices that are not sampled.  The sampled network can thus be roughly regarded as the whole network. 

Both networks are huge and have very high values of the clustering coefficient. Detecting communities accurately within reasonable time in such huge networks is a great challenge for many detection algorithms. PCMA, however, successfully detected millions of communities in both networks within $1$ or $2$ days, using only several ordinary workstations. The details of the parameters used in detection are given in Appendixes~\ref{sec:appendix_1} and~\ref{sec:appendix_2}.

\begin{table*}[htb]
    \centering
    \caption{Information of the two analysed huge social networks}
    \label{tab:social_networks_analyzed}
    \begin{tabular*}{\linewidth}{@{\extracolsep{\fill}}l c c c c c c c}
        \hline
        \ Dataset  &  $n$  &  $m$  &  $\left<k\right>$  &  $C_{\mathrm{WS}}$  &  $c$  &  $T_{1}$  & $T_{2+3}$  \\
        \hline
        \ Friendster  &  65.6M  &   1806M  &  55.1  &  0.205  & 1.6M  &  24.9h / node $\times$ 7 nodes  &  14.2h\\
        \ Sina Weibo  &  79.4M  &  1046M   &  26.4  &  0.155  &  1.2M  &  17.5h / node $\times$ 7 nodes  &  5.1h  \\
        \hline
    \end{tabular*}\par\smallskip
    \raggedright
M stands for million, h stands for hour. $n$, $m$, $\left<k\right>$, and $C_{\mathrm{WS}}$ are the number of vertices, the number of edges, average degree, and the average local clustering coefficient, respectively. $c$ is the number of communities detected by PCMA. $T_1$ and $T_{2+3}$ are the time spent for PCMA Step 1 and Step 2 \& 3, respectively. Step 1 was set to repeat $r=10$ times and pick the best results. It was parallelized on $7$ computing nodes with each equipped with $2 \times$Intel Xeon E5-2670 @ 2.60GHz. This step could be $10$ times faster if $r=1$ is used, just as we did in benchmarking.
\end{table*}

\begin{figure*}[htbp]
   \centering
   \epsfig{file=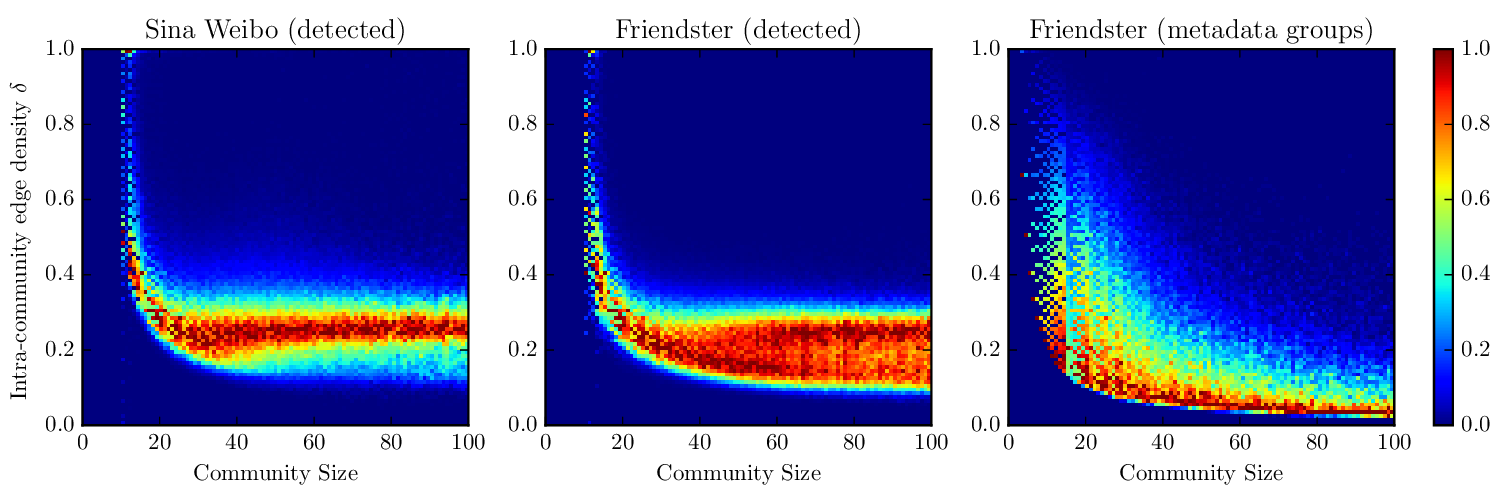, width=\linewidth}
   \caption{(Color online) Intra-community edge density $\delta$ is used to evaluate the quality of a community. Histograms of $\delta$ and community size are shown for the detected communities with size no less than $t_l$ in both social networks and the metadata groups of Friendster. To show the patterns clearly, the values in each vertical cut are rescaled by mapping the highest value to unity. The $0.96$ million metadata groups are of low quality compared to the $1.6$ million communities detected by PCMA in Friendster.}
   \label{fig:intra_edge_density}
\end{figure*}

\subsection{Detected Structural Communities v.s. Metadata Groups}

In this subsection, we show that the millions of communities detected by PCMA in both networks are of high quality, and the metadata groups should not be treated as the ground-truth of the structural communities, of which the definition is purely based on network topology.

\begin{figure}[htbp]
   \centering
   \epsfig{file=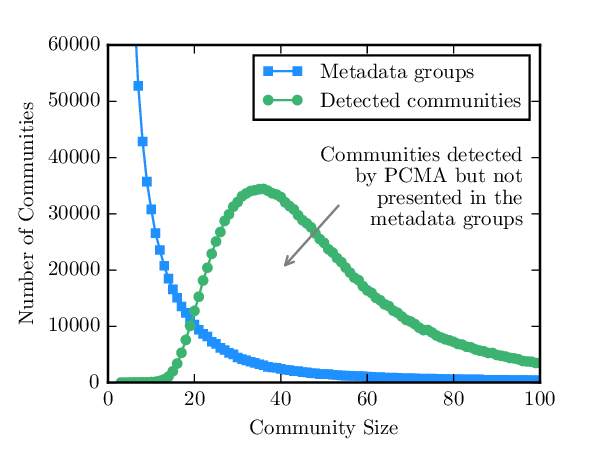, width=0.7\linewidth}
   \caption{(Color online) The size distribution of the metadata groups and the structural communities detected by PCMA in Friendster dataset. The first $4$ data points of the metadata groups staring from $(3, 209764)$ are beyond the shown area of the figure.}
   \label{fig:community_size_dist}
\end{figure}

The number of metadata groups provided in Friendster dataset is about $0.96$ million, while PCMA detected a lot more, $1.6$ million communities. We calculated the Normalized Mutual Information (NMI) between them and the result was approximately $0.004$. It means either PCMA failed to detect the communities or the metadata groups are not the true ground-truth of the structural communities we aim to detect. To figure it out, we adopted the intra-community edge density $\delta$ to evaluate the quality of the detected communities as well as the metadata groups. The density $\delta$ is defined as the number of edges within a community over the maximum number of possible edges there could be. Fig.~\ref{fig:intra_edge_density} shows the results for communities detected in both networks, and the metadata groups of Friendster. Surprisingly, we find that the majority of the metadata groups have very low quality. For example, a group of size $20$ with $\delta=0.1$ means each member knows only $1.9$ other members on average. The group could not even be connected. On the contrary, PCMA successfully controlled the quality of the detected communities in both networks, as we can see that the densities of these communities are reasonably high and seem no lower than some specific value. This successfulness is rooted in the design of PCMA. Recall that $g(C)$ introduced in Sec.~\ref{subsec:merger} can roughly be regarded as $\delta$ and it is guaranteed to be no less than the threshold $t_{f_s}$ used in PCMA Step 2. That means $t_{f_s}$ effectively sets a soft lower bound of the intra-community edge density, i.e. $\delta \approx g \geqslant t_{f_s}$. In this analysis, we used $t_{f_s}=0.1$. Fig.~\ref{fig:intra_edge_density} also shows that some of the metadata groups are of high quality, but they seem not in the same place of the plot of the detected communities. One explanation is that these high quality groups are cores of communities. PCMA does not only find the core members of a community, but may also include peripheral members as long as $\delta$ remains relatively high compared to the given $t_{f_s}$. To verify this speculation, we screened about $0.23$ million high quality groups with $s \geqslant 10$ and $\delta > 1/\sqrt{s}$, where $s$ is the group size, then calculated the recall score for each of them
 \begin{equation}
   R(G) = \max_{C \in \mathscr{D}} \frac{G \cap C}{|G|} \;,
  \label{eq:recall}
\end{equation}
where $G$ represents one of the screened metadata groups, $\mathscr{D}$ is the set of the detected communities of Friendster. In contrast to the very low NMI, the average recall score $\left<R\right>$ reached $0.7$, meaning the detected communities already included on average $70\%$ of each of the screened high quality groups. In fact, PCMA not only included these groups, but also detected a lot more high quality communities that are not presented in the metadata groups, as we can see in Fig.~\ref{fig:community_size_dist} that the number of detected communities of size above $20$ is much greater than the number of metadata groups. We checked carefully whether some of the detected communities were duplicate by calculating the overlap between any pair of these communities. Only a tiny fraction of the pairs have overlap. Among them $86\%$ and $8\%$ of the overlaps are just $1$ or $2$ vertices, respectively. Thus the $1.6$ million communities detected by PCMA are considered authentic and unique.

We conclude that most of the metadata groups of Friendster are of low quality or too small size, while PCMA detected a lot high quality communities that are not in them. The metadata groups may be taken as a reference, but not the ground-truth of the structural communities.

\subsection{Significant Overlaps in Social Communities}

\begin{figure}[htbp]
   \centering
   \epsfig{file=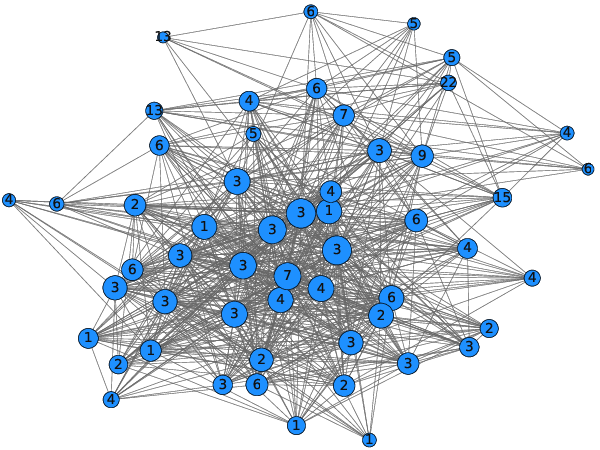, width=0.8\linewidth}
   \caption{A real community consisting of 54 members identified by PCMA in Sina Weibo friendship network. It is the corresponding complete community of the purple partial community shown in Fig.~\ref{fig:first_neighbourhood}. The label on a vertex gives the number of different communities that the vertex belongs to. The number of reciprocal intra-community edges (counted twice) is $702\times2$, and the number of ex-community edges (not shown) connecting to the outside is 5139.}
   \label{fig:c_so}
\end{figure}

\begin{figure}[htbp]
   \centering
   \epsfig{file=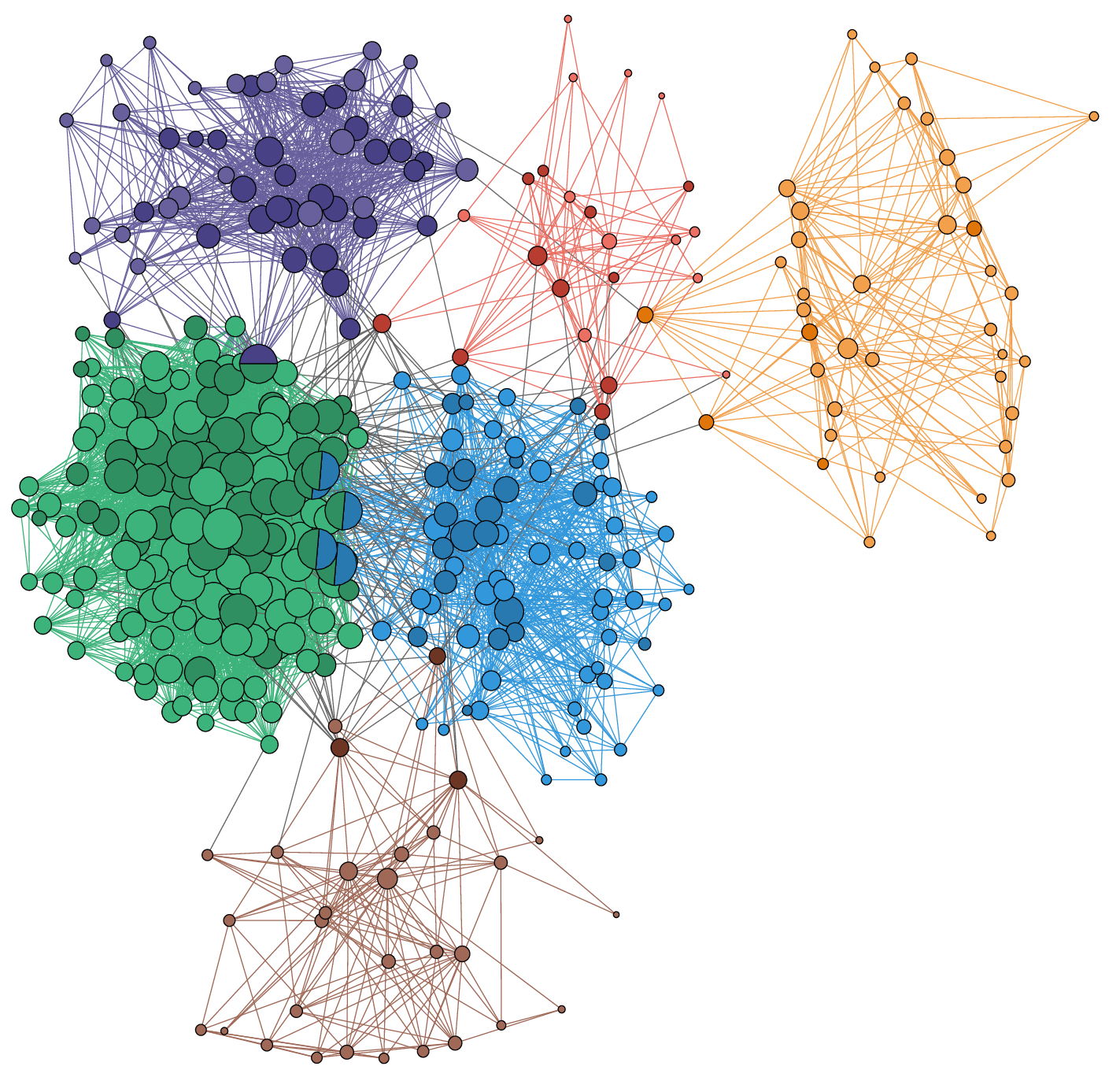, width=0.9\linewidth}
   \caption{(Color online) The communities that the center vertex in Fig.~\ref{fig:first_neighbourhood} belongs to. For clarity, the vertex itself and the edges to its neighbors (vertices with deeper colors) are not shown. The vertex is a core member of $3$ communities (blue, green, purple) as it is connected to a considerable number of their members. These deeper color members are just the $3$ partial communities shown in Fig.~\ref{fig:first_neighbourhood} in the same color. PCMA not only reconstructed the complete communities of these $3$ partial communities, but also found another $3$ that cannot be seen in Fig.~\ref{fig:first_neighbourhood} of which the vertex is a peripheral member. The positions of the vertices calculated by an independent algorithm are in good agreement with the colors determined by PCMA.}
   \label{fig:vertex_communities}
\end{figure}

The detected communities also confirm the multiple memberships of vertices and thus the significant overlaps among communities. Fig.~\ref{fig:c_so} gives an example of a complete community that is only partially revealed in the ego network in Fig.~\ref{fig:first_neighbourhood}. The number on a vertex in Fig.~\ref{fig:c_so} gives the number of communities that the vertex belongs to. To illustrate the multiple memberships of vertices, we picked a vertex labeled $6$ (the same vertex centered in Fig.~\ref{fig:first_neighbourhood}) and visualized the $6$ communities it belongs to in Fig.~\ref{fig:vertex_communities}. For clarity, the vertex itself and the edges to its neighbors (vertices with deeper colors) are not shown. The purple community is the same community shown in Fig.~\ref{fig:c_so}. It overlaps with the other $5$ communities through the picked vertex (not shown). In other words, the vertex acts as a bridge that connects its $6$ communities. Most members of the community, just like the picked vertex, have multiple memberships. They make the community overlap with nearly $200$ other communities while its size is only $54$, resulting in the number of edges going out of the community (ex-community edges) $5139$ to be much larger than the number of reciprocal edges within the community (intra-community edges) $702 \times 2$. It is not an exceptional case,  in fact more than $99\%$ of the communities detected in both social networks are found to have more ex-community than intra-community edges. This confirms that detection algorithms based on concepts for disjoint or slightly overlapping communities do not work in social networks, and new algorithms that can handle multiple memberships of vertices, such as PCMA, are called.

\section{Summary}
\label{sec:summary} We proposed and implemented a Partial
Community Merger Algorithm specifically designed for detecting
communities in large-scale networks in which a member may have multiple
memberships.  The structure of these communities is signified by
a large number of overlaps in members and thus a community may have many more edges
connecting to the outside compared to those within the community.
Such structures make many existing community detection algorithm
invalid, but yet they often show up in real-world systems. Through
PCMA, we provided a conceptual framework as well as a practical
algorithm and significantly advanced the techniques in dealing with these systems. Details of implementing PCMA were discussed.  We used two benchmark models
and compared results with four algorithms to establish
the efficiency and accuracy of PCMA. We also tested the method on two huge social networks and millions of high quality communities were detected. The metadata groups are of low quality compared to the detected and we showed that they should not be considered as the ground-truth of the structural communities that detection algorithms usually aim to detect. We also verified that the multiple memberships of vertices and significant overlaps are extremely common in the detected communities. Thus new algorithms that can detect communities with these complex structures, such as PCMA, are essential and urgently needed.

PCMA does not need a
prior knowledge of the total number of communities to search for and it
is capable of analyzing communities in large-scale networks in
linear time.  In addition to identifying the communities, PCMA also
gives who the key members are in a community and how many
different communities a member belongs to.  The high accuracy and
linear time complexity makes PCMA a promising tool for detecting
communities with significant overlaps in huge social networks,
which cannot be handled by most existing algorithms.

We end with a few remarks.  Although we described and implemented
PCMA only for unweighted networks, the approach is flexible and it
can be readily extended to treat weighted networks. Another
extension is to properly tune the threshold $t_{f_{s}}$ for
exploring the hierarchical structure of communities. Like any
other algorithm, PCMA also has its limitations.  It is not
designed to detect small communities and it will not work in
networks that are too sparse.  There is also the common problem
among algorithms on distinguishing real communities from false
ones.  This is actually a deeper question because whether there
really exists a clear boundary, i.e. a set of thresholds, for distinguishing ``real
communities" from ``false ones" is questionable.  A more
practically approach would be to explore methods of choosing the
thresholds properly or constructing a credibility function involving $g$, $l$,
and community size to make the detected communities more trustful.
These quantities in the PCMA framework provide us with vast space for further improvement in post-processing and in-depth analysis of community structure.

Finally, the source code of PCMA is released
as free software under the GNU General Public License version 2 or
any later version~\cite{Xu:2016ab}.

\section*{Acknowledgements}
One of us (EHWX) gratefully acknowledges the support from the
Research Grants Council of Hong Kong through a Hong Kong PhD Fellowship.

We thank an anonymous reviewer for informative suggestions on related work in the computer science literature.

\appendix
\section{Searching for Partial Communities}
\label{sec:appendix_1}

PCMA Step 1 is to search for partial
communities in the ego network of each vertex.  We adopt an
efficient community detection algorithm proposed by Ball
\textit{et al.}~\cite{Ball:2011wz} for the purpose.  Starting from
an ego network of size $n$, the algorithm takes the number $K$ of
communities to be found as an input. The output is an $n \times K$
matrix, with the $K$ numbers in a row that signifies
the belonging coefficients~\cite{Gregory:2011jv} of a
vertex for the $K$ communities.  For example, we set out to find
$K=5$ communities in an ego network, labelled by $C1$ to $C5$. Then
each row has $5$ numbers, e.g. $0.64$, $0.29$, $0$, $0.07$, $0$
for the $j$-th row, denoting that the vertex $j$ has a portion of
$64\%$ belonging to $C1$, $29\%$ to $C2$, and $7\%$ to $C4$. To
convert the fuzzy assignment~\cite{Gregory:2011jv} of members to
definite assignment, we impose a threshold that a community
carries a vertex $j$ as its member only if the belonging
coefficient of the vertex $j$ for the community is above a certain value.
If the threshold is $0.2$, only communities $C1$ and $C2$ have the vertex $j$
as a member. For the central vertex of the ego network, as the
communities are its partial communities, it is treated as a
default member of all the communities regardless of the threshold.
Partial communities can then be derived from every ego network. We
remark that although the number of partial communities in an ego
network is not known in prior to the search, it can easily be
estimated for small networks.  An appropriate overestimation is
necessary as isolated vertices in an ego network also need to
be accommodated by some communities.  Such overestimation is
harmless as the false partial communities can be handled properly
in the merger and post-processing steps that follow.  It is
reasonable to assume the number of partial communities to be
proportional to the ego network size, optionally with an upper and a lower bound. The overestimation will not cause problem, as the true
communities can be merged when they have a considerable overlap.
For the example in Fig.~\ref{fig:first_neighbourhood}, we set out
to find $10$ communities and obtained $3$ apparent partial
communities (colored purple, blue and green) and $2$ possible
ones (red and yellow). The green one actually consists of $3$
highly overlapping communities which should be regarded as one. We
merge a pair of partial communities if either one shares more than
a certain fraction of the members of the other.

It is important to note that PCMA does \textit{not} require a high
accuracy in this step of finding all the partial communities. Any
error generated in this step can be greatly reduced by the mergers
and post-processing step, as discussed in
Sec.~\ref{subsec:post-processing}.
For example, the detection algorithm we used in this step needs to try $r=10$ to $100$ random initial conditions to get relatively good results, as suggested by the authors of the algorithm~\cite{Ball:2011wz}. We used only one in benchmarking, yet PCMA still performed very well. Increasing the number of random initial conditions could get even better results, although with the improvement not be much. For empirical analysis, we used $r=10$ to get more accurate results.
The parameters used in empirical analysis are summarized in Table~\ref{tab:parameters_step_1_e}. For benchmarking, $K=5$, , $r=1$, $b=0.2$, $m=0.3$. Parameter $c$ and $d$ are not used. 

\begin{sidewaystable}
    \centering
    \caption{Parameters of PCMA Step 1 used in the empirical analysis}
    \label{tab:parameters_step_1_e}
    \begin{tabular*}{\linewidth}{@{\extracolsep{\fill}} c c l }
        \hline
        Parameter	&	Value		&	Description	\\
        \hline
        \multirow{2}{*}{$K$}		&	\multirow{2}{*}{-}	&	\multirow{2}{*}{\parbox{15cm}{\raggedright The number of partial communities set to be found in a vertex's ego network is $K=\mathrm{min} \{3 + k/10, 20 \}$, where $k$ is the degree of the vertex}}	\\
        && \\
        \multirow{1}{*}{$r$}		&	\multirow{1}{*}{$10$}		&	\multirow{1}{*}{Repeat the detection of partial communities for $r$ times, pick the best result}	\\

        \multirow{1}{*}{$b$}		&	\multirow{1}{*}{$0.2$}	&	\multirow{1}{*}{A vertex is regarded as a member of a partial community if its belonging coefficient to this community is above $b$}	\\

        \multirow{1}{*}{$m$}		&	\multirow{1}{*}{$0.2$}	&	\multirow{1}{*}{Merge partial communities of the same vertex if the overlap between two partial communities is above the threshold}	\\
	\hline
        \multirow{1}{*}{$c$}		&	\multirow{1}{*}{$0.05$}	&	\multirow{1}{*}{Discard partial communities of which the clustering coefficient (treat the community as a subnetwork) is below $c$}	\\

        \multirow{1}{*}{$d$}		&	\multirow{1}{*}{$10$}		&	\multirow{1}{*}{Skip  vertices with degree lower than $d$ as it is less likely to find partial communities in these vertices' ego networks}	\\
        \hline
    \end{tabular*}

    \vspace{4\baselineskip}
    
    \caption{Thresholds used in PCMA Step 2 \& 3 for both benchmarking and empirical analysis}
    \label{tab:threshold_pcma_2_3}
    \begin{tabular*}{\linewidth}{@{\extracolsep{\fill}} c c l }
        \hline
        Threshold & Value &  Description \\
        \hline
        \multirow{1}{*}{$t_{f_s}$}	&	 \multirow{1}{*}{$0.1$}	&	\multirow{1}{*}{\parbox{15cm}
        {\raggedright A pair of partial communities can be merged only if the similarity $f_s$ between them is above $t_{f_s}$}}	\\
        \multirow{1}{*}{$t_{f_0}$}	&	 \multirow{1}{*}{4}	&	\multirow{1}{*}{Used to suppress unwanted mergers of small partial communities}	\\
        \hline
        \multirow{1}{*}{$t_{l}$}	& 	\multirow{1}{*}{10}	&	\multirow{1}{*}{A merged community must contain at least $t_l$ partial communities to be regarded as a real complete community}	\\
        \multirow{1}{*}{$t_{S'/l}$}	&	\multirow{1}{*}{0.1}	&	\multirow{1}{*}{The score $S'$ of a member (its own contribution excluded) must be greater than $t_{S'/l}\cdot l$}\\
        \multirow{1}{*}{$t_{S'}$}	&	\multirow{1}{*}{2}	&	\multirow{1}{*}{The score $S'$ of a member (its own contribution excluded) must be no less than $t_{S'}$}	\\
        \hline
    \end{tabular*}
\end{sidewaystable}

\begin{figure}
   \centering
   \epsfig{file=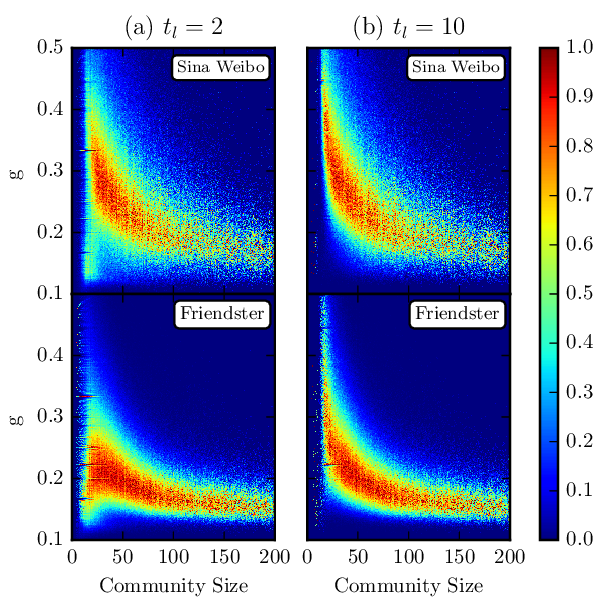, width=0.7\linewidth}
   \caption{(Color online) The histogram of communities as a function of their value of $g$ and community size provides an alternative to inspect the quality of the detected communities. Each vertical cut gives the distribution of $g$ for communities of the same size. The values in each cut are rescaled by mapping the highest value to unity.}
   \label{fig:g_2d}
\end{figure}

\begin{figure}[htbp]
   \centering
   \epsfig{file=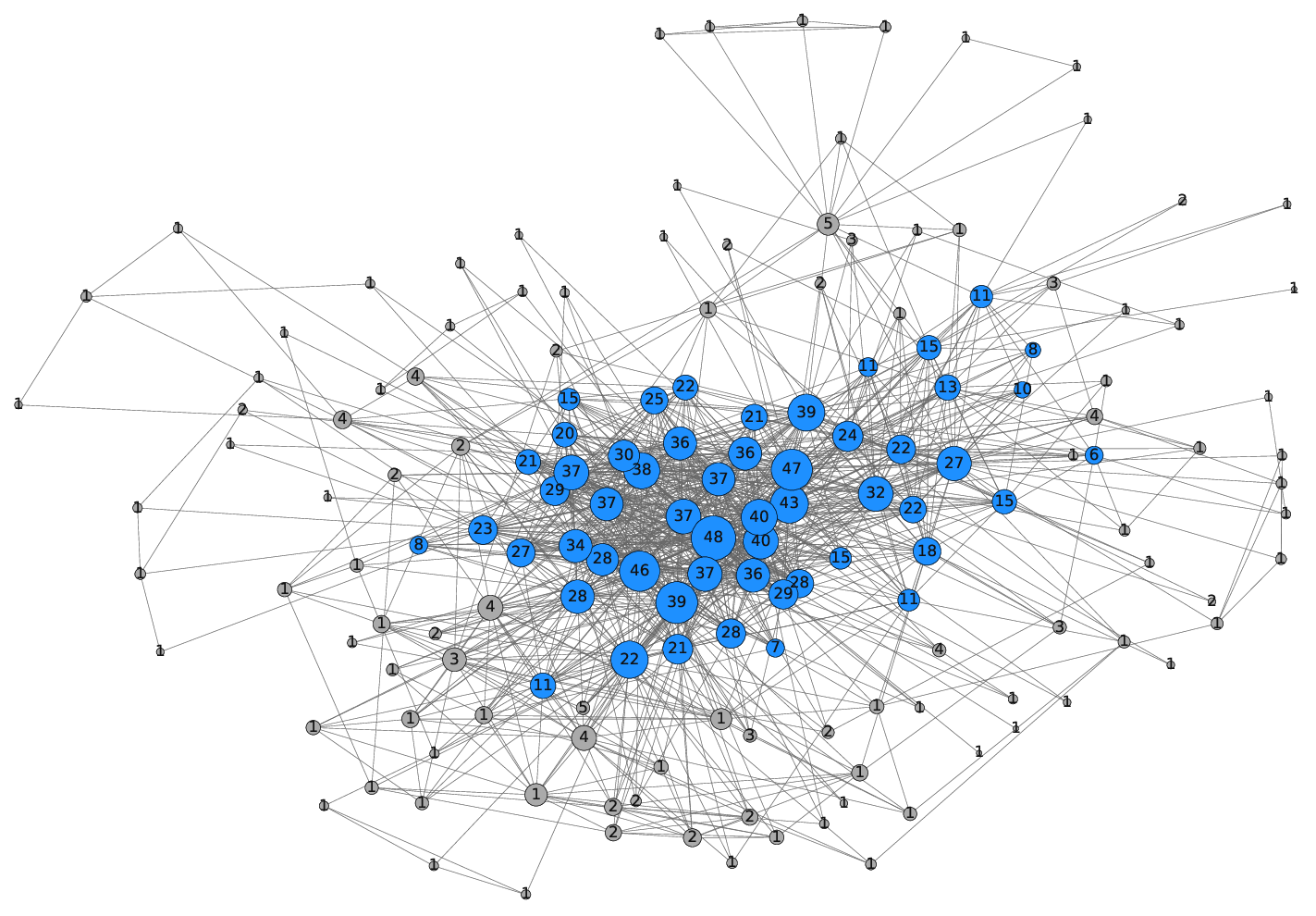, width=\linewidth}
   \caption{(Color online) A merged community with false members not removed. It is merged from $l=54$ partial communities and has $160$ vertices before post-processing. The size (area) of each vertex represents its degree in this subnetwork. The label on each vertex (zoom in to see the numbers) gives the score $S'$ of the vertex. The grey vertices with $S' < 6 $ fail to meet the threshold $t_{S'/l}=0.1$ so they are not considered members. The remaining $54$ blue vertices are members of the community.}
   \label{fig:c_before_step_3}
\end{figure}

\section{Thresholds in PCMA Step 2 \& 3}
\label{sec:appendix_2}

The thresholds used in PCMA Step 2 \& 3 for both benchmarking and empirical analysis are summarized in Table~\ref{tab:threshold_pcma_2_3}. For more information about the physical meanings and proper use of the thresholds, see Sec.~\ref{subsec:post-processing} and Sec.~\ref{subsec:merger}. Here we briefly illustrate how the applied thresholds remove false communities and false members.

Fig.~\ref{fig:g_2d} shows the histograms for the detected communities as a function of their value of $g$ and community size, under two different thresholds $t_l=10$ and $2$. The quantity $g$ is another community quality indicator in addition to the intra-community edge density $\delta$. Usually a smaller community would have a higher $g$. Choosing a low threshold of $t_{l}=2$ gives an abnormal plunge in $g$ at small community sizes as shown in Fig.~\ref{fig:g_2d}(a). The low threshold leads to many false communities that are merged only a few times, as discussed in Sec.~\ref{subsec:post-processing}.  Raising $t_{l}$ from $2$ to $10$ successfully removes most false communities, as shown in Fig.~\ref{fig:g_2d}(b). And the number of communities detected in Friendster and Sina Weibo is reduced from $9.0$ million to $1.6$ million, $4.7$ million to $1.2$ million, respectively. This success is accompanied by the drawback that real communities with less than $t_l=10$ vertices are also removed. Note that there is no best answer to the value of $t_l$ because the boundary between true and false communities itself is indistinct. A more practically approach is to find a function involving $g$, $l$, and community size to evaluate the credibility of a detected community, instead of labelling them either true or false. We leave it to future work.

For $t_{S'/l}$ and $t_{S'}$, there is no standard answer to which values are the best either. It depends on the network being analyzed, and one's subjective view of what a community is. Fig.~\ref{fig:c_before_step_3} illustrates a merged community that is the same community shown in Fig.~\ref{fig:c_so} but with false members not removed. Since $l=54$ and $t_{S'/l}=0.1$, only vertices with $S' \geqslant 6$ are considered members (colored blue). As we can see in Fig.~\ref{fig:c_before_step_3}, the thresholds successfully sift out the core of the community. It is, however, hard to say whether vertices in the periphery with $S'=5$ or $4$ are members or not. In fact, there does not exist a clear boundary of the community. A better solution is to assign each vertex a belonging coefficient representing its degree of affinity to the community, e.g. $S'/l$, instead of forcing each vertex to be either in or not in the community.

The quantities $S'$, $l$, $g$, and $f_s$ introduced in the PCMA framework have sound physical meanings and they provide us with vast space for further improvement in post-processing and in-depth analysis of community structure.

\bibliographystyle{nws}
\bibliography{pcma}

\end{document}